# Automated Generation of Interorganizational Disaster Response Networks through Information Extraction

**Yitong Li**
George Mason University
yli63@gmu.edu

**Duoduo Liao**
George Mason University
dliao2@gmu.edu

**Jundong Li**
University of Virginia
jl6qk@virginia.edu

**Wenying Ji**
George Mason University
wji2@gmu.edu

**ABSTRACT**

When a disaster occurs, maintaining and restoring community lifelines subsequently require collective efforts from various stakeholders. Aiming at reducing the efforts associated with generating Stakeholder Collaboration Networks (SCNs), this paper proposes a systematic approach to reliable information extraction for stakeholder collaboration and automated network generation. Specifically, stakeholders and their interactions are extracted from texts through Named Entity Recognition (NER), one of the techniques in natural language processing. Once extracted, the collaboration information is transformed into structured datasets to generate the SCNs automatically. A case study of stakeholder collaboration during Hurricane Harvey was investigated and it had demonstrated the feasibility and applicability of the proposed method. Hence, the proposed approach was proved to significantly reduce practitioners' interpretation and data collection workloads. In the end, discussions and future work are provided.

**INTRODUCTION**

Disasters often cause devastating impacts on community lifelines (Homeland Security, 2016, 2019). When a disaster occurs, maintaining and restoring community lifelines heavily rely on collective efforts, which a single stakeholder could hardly accomplish (National Research Academies, 2012; Nowell et al., 2018). Large-scale disasters (e.g., hurricanes, earthquakes, fires, and pandemics) always involve a great number of stakeholders, such as multi-level (i.e., federal, state, and local) governments, nongovernmental organizations, and private sectors, which are with various roles, responsibilities, and interactions, thereby increasing the complexity of response collaborations (Homeland Security, 2019). To incorporate complex interactions among stakeholders, network analysis has been widely applied by modeling stakeholders and their interactions as nodes and edges, respectively (Comfort, 2007; Djalante et al., 2013; Nowell et al., 2018; Li and Ji, 2020). Using the modeled network, quantitative analyses are further conducted to identify high-performing stakeholders or groups of stakeholders (Kapucu, 2005; Kapucu and Garayev, 2013; Nowell et al., 2018).

To generate stakeholder collaboration networks (SCNs), collaboration information that identifies stakeholders and stakeholder interactions is needed. Nowadays, various types of documents are provided to 1) summarize stakeholders' roles and responsibilities (Homeland Security, 2019; Federal Emergency Management Agency, 2019) for guiding stakeholder collaboration and 2) track actual stakeholder response actions for updating information (Federal Emergency Management Agency, 2017; The Texas Division of Emergency Management, 2020; Federal Emergency Management Agency, 2020; Chen et al., 2020). These documents comprise valuable collaboration information to identify stakeholders and understand their interactions. Although existing in various documents, extracting SCN components (i.e., stakeholders and stakeholder interactions) is challenging because the majority of collaboration information is embedded in unstructured texts. When the scale of a disaster becomes large, the number of documents increases, thereby making the information extraction process more labor-intensive and time-consuming. In addition, the subjectivity of human interpretations impedes the consistency of the extracted information, which, in turn, affects the accuracy of the generated SCNs. Therefore, to reduce practitioners' efforts in manually reviewing text documents, and to ensure the consistency of the extracted information, a systematic approach that is capable of automatically generating SCNs from texts is highly desired.

The objective of this research is to propose a systematic approach for reliably extracting stakeholder collaboration information from texts and automatically generating SCNs. The core of the proposed approach is a Natural Language Processing (NLP) technique—Named Entity Recognition (NER), which functions to automatically identify stakeholders from texts. Once identified, stakeholder interactions are established if stakeholders are involved in the same response tasks (e.g., debris removal, hazardous waste collection, and damage assessment). Lastly, stakeholders and stakeholder interactions are transformed into two structured datasets to automate the generation of SCNs. Overall, the proposed approach achieves the reliable and automated generation of SCNs from texts, which largely reduces practitioners' interpretation loads and eases the data collection process. The remainder of this paper is organized as follows. In the next section, motivations of using NER to obtain collaboration information are discussed. After that, a systematic approach designed for achieving the automated generation of SCNs is explained step by step. Following the methodology section, federal stakeholder collaboration in response to Hurricane Harvey was analyzed to demonstrate the feasibility and applicability of the proposed approach. In the end, contributions and future work are concluded.

**RATIONALE**

In the disaster management domain, network analysis has been widely applied to study complex interactions among stakeholders (Kapucu, 2005; Kapucu and Garayev, 2013; Nowell et al., 2018). To obtain collaboration information for generating SCNs, practitioners widely rely on conducting surveys and consulting with agencies who have navigated the disaster response process (Bodin and Nohrstedt, 2016; Li et al., 2019, Homeland Security, 2019). Although surveys are a very useful tool to gather collaboration information, designing surveys and arranging meetings with various agencies require a large amount of time and high coordination efforts (Jones et al., 2013). Most importantly, it is difficult to ensure the objectivity of the collected information due to the limited access to response agencies (Spence and Lachlan, 2010). Nowadays, collaboration information on disaster response has been recorded in various government-issued documents to guide disaster response and track stakeholder response actions (Federal Emergency Management Agency, 2019, 2020). Compared to surveys, these documents are easy to access and have integrated collaboration views among various agencies (i.e., more objective), which makes them a promising data source for obtaining collaboration information. Specifically, an example from the firefighting ESF Annex is "*the **Department of the Interior** assists the **U.S. Department of Agriculture/Forest Service** in managing and coordinating firefighting operations* (Homeland Security, 2019)." In this example, the involved stakeholders are Department of the Interior and U.S. Department of Agriculture/Forest Service, and these two stakeholders interact with each other. This piece of collaboration information can be represented in a network format—two nodes (i.e., Department of the Interior and U.S. Department of Agriculture/Forest Service) are linked by one edge (i.e., Department of the Interior interacts with U.S. Department of Agriculture/Forest Service). As more collaboration information becoming available, comprehensive SCNs can be generated for analyzing stakeholder collaboration during disaster response.

Although these documents contain rich collaboration information, the information is mostly embedded in texts. To reduce human efforts associated with reviewing these documents, and to reduce the subjectivity of human interpretations, a process that can automatically extract the collaboration information is desired. Information extraction (IE) is a process that turns the unstructured information embedded in texts into structured data (Jurafsky and Martin, 2019). Among various IE tasks (e.g., NER, relation extraction, and event extraction), NER is an essential one, which seeks to locate and classify named entities mentioned in texts into pre-defined categories, such as person names, organizations, and locations (Jurafsky and Martin, 2019). In detail, NER deals with deciding whether a span of texts is an entity or not, and where the boundaries are. Once all named entities have been extracted, entities that refer to the same real-world entities are linked. For example, the Federal Emergency Management Agency and FEMA refer to the same disaster response agency. In the context of the presented research, NER is capable of automatically processing texts, extracting stakeholders as organization entities, and linking entities that refer to the same stakeholders. Therefore, NER is used to extract stakeholders for generating SCNs.

**METHODOLOGY**

To achieve the objective of automatically generating SCNs from unstructured texts, a systematic approach is proposed and shown as Figure 1. First, text documents that contain stakeholder collaboration information related to disaster response are selected. Then, a dictionary that contains the names of interested stakeholders is specified. Based on the specified dictionary, NER is applied to automatically extract stakeholders. Once extracted, the information is transformed into two structured datasets (e.g., stakeholders and stakeholder interactions) to generate SCNs.

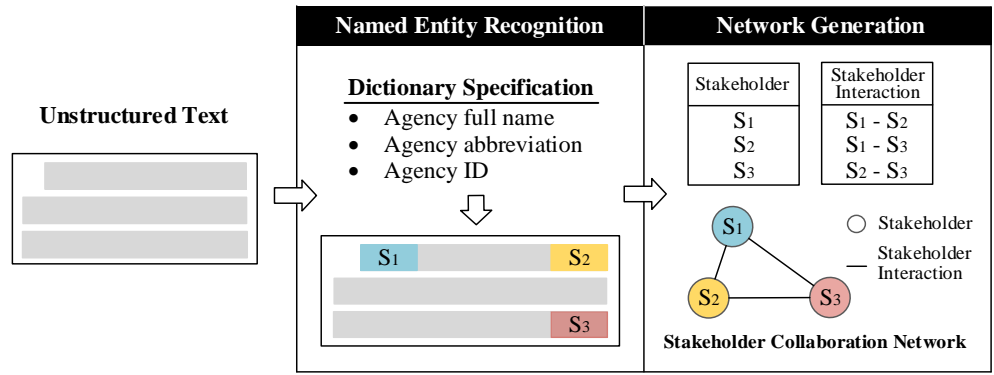

**Figure 1. Research Framework**

### Named Entity Recognition

To extract stakeholders from the selected text documents, a dictionary that contains the names of interested stakeholders is specified first. In this research, stakeholders are defined as government agencies that are responsible for disaster response. Since most agency names are well documented in government-issued guidelines (Federal Emergency Management Agency, 2019, 2020), the dictionary can easily be specified. In the dictionary, both full names and abbreviations of agencies are included because the two formats are widely used in text documents to refer to agencies. To link agency names that refer to same stakeholders, the dictionary includes a unique ID that is assigned to an agency's full name and its abbreviation. For example, the ID "FEMA" is assigned to "Federal Emergency Management Agency" and "FEMA" to indicate that they refer to the same stakeholder.

Once the dictionary is specified, NER is applied to extract stakeholders from the selected text documents. Before applying NER, the text documents and agency names in the specified dictionary are converted to the same letter case representations (i.e., lower case or upper case). This is to incorporate agency names with letter case variations (e.g., "Fema", "fema", and "FEMA"), and to reduce the number of agency names required in the dictionary. Once letter cases are kept consistent, NER is applied to process texts and automatically extracts the interested stakeholders.

### Network Generation

To generate SCNs, two types of information are needed: stakeholders (i.e., nodes) and stakeholder interactions (i.e., undirected edges). Stakeholders are extracted from texts using NER. Stakeholders are extracted from texts using NER. Based on the extracted stakeholders, stakeholder interactions are obtained by establishing pairwise interactions when these stakeholders are involved in the same response tasks. For example, if three stakeholders (e.g., S1, S2, and S3) are involved in one response task (e.g., debris removal, hazardous waste collection, and damage assessment), then there are three types of stakeholder interactions (e.g., S1-S2, S1-S3, and S2-S3). To automate the network generation, stakeholders and stakeholder interactions are transformed into two structured datasets, namely, a node dataset and an edge dataset. By importing the datasets into network analysis software (e.g., igraph [Csardi and Nepusz, 2019]), SCNs are automatically generated. Once generated, network analysis can be conducted to evaluate the performance of stakeholder collaboration.

### CASE STUDY

In this section, a SCN of federal stakeholders involved in Hurricane Harvey is generated to demonstrate the feasibility and applicability of the proposed approach. Hurricane Harvey was a category 4 hurricane that caused devastating damages to southeastern Texas. It is the second-most costly hurricane in U.S. history, and has the largest number of direct deaths in Texas since 1919 (Blake and Zelinsky, 2018). During disaster response, most state and local stakeholders were overwhelmed by disaster impacts and required assistance from federal stakeholders, which makes the role of federal stakeholders critical (Homeland Security, 2019). In this research, a 10-day period starting from the date when Hurricane Harvey made landfall (Aug. 26, 2017) to the date when Harvey's impact started dissipating (Sept. 04, 2017) is selected. The case study is conducted in the following steps. First, data sources used to extract stakeholder collaboration information are selected. Then, the proposed methodology is applied to identify stakeholders involved during disaster response. After this, the extracted stakeholders are transformed into structured datasets for the network generation. To demonstrate the usage of the SCN, individual stakeholders' performances are compared using three centrality measures (i.e., degree centrality, closeness centrality, and betweenness centrality). Lastly, two perspectives of SCN analyses are summarized for enhancing stakeholder collaboration in disaster response.

**Data Source**

Mission Assignment (MA) provided by FEMA (Federal Emergency Management Agency, 2020) was selected as the data source. In detail, MA is a work order issued by FEMA to other Federal agencies when state and local governments are overwhelmed by disaster impacts. In the MA dataset, each MA includes essential information required to direct the completion of a specified response task (e.g., debris removal, hazardous waste collection, and damage assessment), such as cost estimation, location of need, statement of work (SOW), and assigned agency. Among all types of information, agency and SOW were selected. In detail, agency contains names of agencies receiving the MA from FEMA, which can be used to specify the stakeholder dictionary. SOW contains text descriptions of how stakeholders should collaborate in providing the required assistance, which is selected as text descriptions to extract collaboration information. Since each SOW corresponds to one response task, pairwise stakeholder interactions are established for stakeholders extracted from the same SOW. After filtering the MA, 294 pieces of SOWs were extracted during the investigated time period (i.e., Aug. 26, 2017 - Sept. 04, 2017). A sample SOW is shown in Figure 2. Using the proposed methodology, the extracted stakeholders are "FEMA" (Federal Emergency Management Agency), "DOE" (Department of Energy), and "NRCC" (National Response Coordination Center). Stakeholder interactions are FEMA-DOE, FEMA-NRCC, and DOE-NRCC. Based on the collaboration information, a SCN is generated and shown on the right of Figure 2.

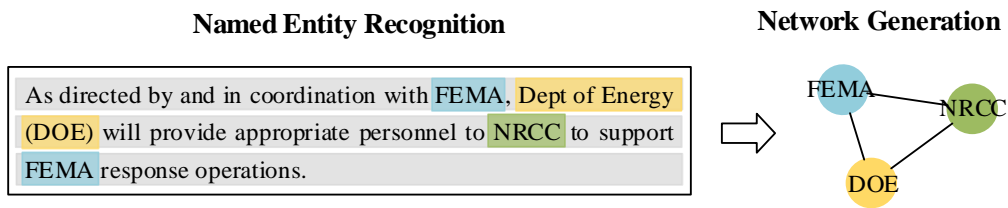

**Figure 2. Illustrative Example**

**Named Entity Recognition**

To extract stakeholders from the selected texts, the stakeholder dictionary was first specified. In detail, full names and abbreviations of agencies given in the MA were included, and unique IDs were assigned to link agency names that refer to the same stakeholders. In the dictionary, 193 agency names were included, and 102 unique IDs were assigned. A sample of the labeled stakeholders is illustrated in Table 1.

**Table 1. Sample Stakeholders and Stakeholder IDs**

| Agency | Stakeholder ID |
| --- | --- |
| Federal Emergency Management Agency | FEMA |
| FEMA | FEMA |
| American Red Cross | ARC |
| ARC | ARC |
| Department of Defense | DoD |
| DoD | DoD |
| Health and Human Services | HHS |
| HHS | HHS |
| Federal Protective Service | FPS |

| FPS | FPS |
|---|---|

In this research, the open-source library spaCy (spaCy, 2020) was used to perform NER. SpaCy was selected because it is easy to implement and it generates accurate outputs in a relatively fast manner. Most importantly, the pre-trained NER model in spaCy can easily be specialized by adjusting the processing pipeline, which enables the specialization of the pre-trained NER model by adding the specified dictionary. To achieve the automated extraction of stakeholders, the EntityRuler function in spaCy was used (spaCy, 2020). In detail, EntityRuler enables the addition of named entities (i.e., agency names) by specifying entity patterns. Here, the entity patterns were specified with three keys: label, which specifies the label to assign to the entity if the pattern is matched; pattern, which is the matched pattern; and ID attribute, which allows multiple patterns to be associated with the same entity. Once entity patterns were specified, EntityRuler finds matched entities in the texts and assign them labels and IDs. In this research, the label assigned to the entity is "stakeholder", entity patterns are agency names shown in Table 1, and ID attributes are stakeholder IDs shown in Table 1. Using the entity patterns, EntityRuler is able to find agency names in texts, label agency names as stakeholders, and assign stakeholder IDs to agency names.

**Stakeholder Collaboration Network Generation**

To generate the SCN, stakeholders and stakeholder interactions are needed. To extract stakeholders, the 294 pieces of SOWs, and agency names in the dictionary were first converted to lower case letters. Then, the specialized NER was applied to extract stakeholders mentioned in each SOW. Once extracted, pairwise interactions of stakeholders obtained from one SOW were established. Following this process, 43 unique stakeholders and 113 unique stakeholder interactions were obtained. Using IDs to represent stakeholders, sample stakeholders and stakeholder interactions are shown in Table 2 and Table 3, respectively.

**Table 2. Sample Stakeholders**

| Stakeholder ID |
|---|
| FEMA |
| DHS |
| NPPD |
| HHS |
| NRCC |
| FPS |
| OPS |
| OHA |
| CNCS |
| DOE |

**Table 3. Sample Stakeholder Interactions**

| Stakeholder ID (a) | Stakeholder ID (b) |
|---|---|
| FEMA | DHS |
| FEMA | NPPD |
| DHS | NPPD |
| FEMA | HHS |
| FEMA | NRCC |
| HHS | NRCC |
| FEMA | FPS |
| NRCC | FPS |
| FEMA | OPS |

To ensure the accuracy of the extracted stakeholders, the authors have randomly selected 100 pieces of SOWs and identified the desired stakeholders. By comparing the total number of extracted stakeholders (244) with the total number of desired stakeholders (262), the specialized NER is able to extract stakeholders with 93.13% accuracy.

The error is mainly due to the incompleteness of the specified dictionary (i.e., certain desired stakeholders are not specified), which can be reduced by enriching the dictionary. Another reason is that stakeholders extracted from texts do not necessarily refer to stakeholders. For example, in the text "FAA regulations and restrictions," the specialized NER identifies FAA (i.e., Federal Aviation Administration) as a stakeholder, whereas FAA refers to a regulation type. To reduce this error, more advanced NLP techniques (e.g., semantic role labeling) that consider text semantics are needed.

Using the extracted collaboration information, the SCN was generated and shown as Figure 3. The network was visualized using network visualization package igraph (Csardi et al., 2019) in R (R Core Team, 2020).

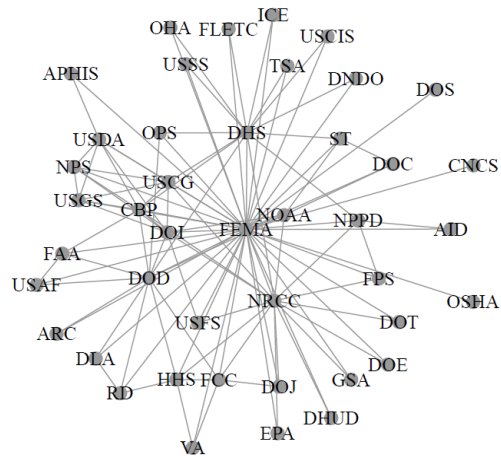

**Figure 3. Federal Stakeholder Collaboration Network**

To demonstrate the usage of the generated SCN, network analysis was conducted by comparing the performance of individual stakeholders using three types of centrality measures (i.e., degree centrality, closeness centrality, and betweenness centrality [Scott, 1988]). Specifically, degree centrality counts the number of links incident upon a node, and stakeholders with high degree centrality scores will have more influence on other stakeholders. Closeness centrality is defined as the inverse of the farness, which measures how close a node is to all other nodes. In the SCN, stakeholders with high closeness centrality scores are able to efficiently reach every other stakeholder in the network. Lastly, betweenness centrality measures the number of shortest paths between pairs of nodes that pass through a node, and it is an indication of stakeholder's ability to control over collaboration between others (Scott, 1988). In Figure 4, the top five stakeholders with high centrality scores were visualized, and stakeholder IDs were labeled on the x-axis.

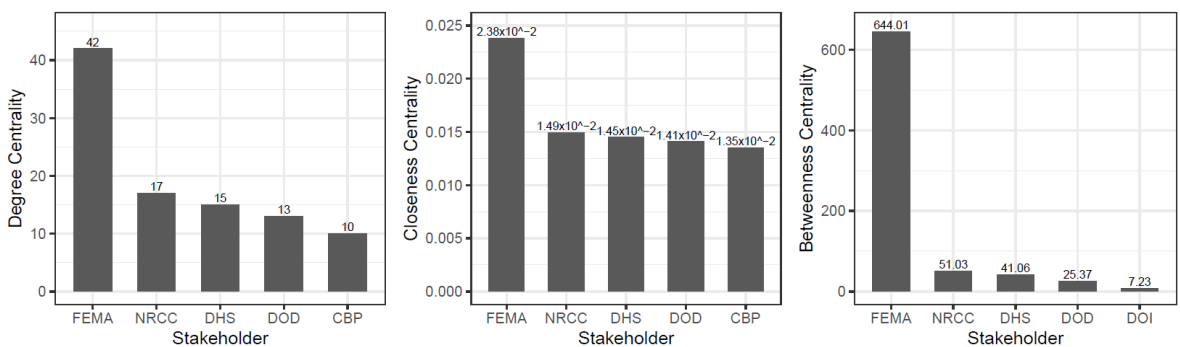

**Figure 4. Top 5 Stakeholders by Centrality Measures**

Overall, six high-performing stakeholders (i.e., FEMA, NRCC, DHS, DOD, CBP, and DOI) were identified. These stakeholders are recommended to share their response experience and lessons learned with other stakeholders. In addition, the top five stakeholders with high degree centrality scores are also stakeholders with high closeness centrality scores, which shows that stakeholders who have high influences on others are stakeholders who can efficiently reach others. This finding indicates that stakeholders' ability to efficiently reach other stakeholders is a critical property of stakeholder collaboration in disaster response.

**DISCUSSION**

To illustrate the applicability of the generated SCN for enhancing stakeholder collaboration in disaster response, two perspectives were summarized: 1) identification of high-performing stakeholders and stakeholder subgroups and 2) analysis of the evolution of stakeholder collaboration. Specifically, high-performing stakeholders and stakeholder subgroups can be identified using quantitative network metrics that measure the performance of individual stakeholders (e.g., degree centrality, closeness centrality, and betweenness centrality [Scott, 1988]) and subgroups of stakeholders (e.g., giant components and cliques [Boccaletti et al., 2006]), respectively. Once identified, exemplary stakeholders should be invited to share their valuable experience and lessons learned from the past disaster response.

To understand the evolution of stakeholder collaboration over time, high-performing stakeholders or collaboration patterns at different timestamps can be examined. Based on the results, time periods that have characteristic changes of collaboration patterns (i.e., emerging stakeholders or/and stakeholder interactions) can be identified and further mapped to actual disruptive events (e.g., food shortage, power outage, highway inundation). Through correlating actual disruptive events with their corresponding collaboration patterns, valuable collaboration strategies can be generalized. Once obtained, these strategies can be used to revise existing disaster response guidelines for enhanced stakeholder collaboration.

**CONCLUSION**

Disasters cause devastating damages to community lifelines. To maintain and restore community lifelines, stakeholders need to collaborate with each other for taking effective response actions. To better understand stakeholder collaboration, network analysis has been widely applied through modeling stakeholders and their interactions as nodes and edges, respectively. Nowadays, various text documents contain rich collaboration information for generating SCNs. To reduce practitioners' efforts in manually reviewing text documents, and to ensure the consistency of the extracted information, a systematic approach that achieves the reliable extraction of collaboration information from texts and the automated generation of SCNs is proposed. In the proposed approach, NER is applied to automatically extract stakeholders from texts. Once extracted, stakeholder interactions are established if they are mentioned in the same response tasks. Once stakeholders and stakeholder interactions are identified, these types of information are transformed into structured datasets for automated generation of SCNs. A case study on federal stakeholder collaboration in response to Hurricane Harvey was analyzed to demonstrate the feasibility and applicability of the proposed approach. From the case study, six high-performing federal stakeholders were identified, and these stakeholders are recommended to share their response experience for improved stakeholder collaboration. Also, it is observed that stakeholders' ability to efficiently reach other stakeholders is a critical property of stakeholder collaboration in disaster response. Lastly, two perspectives of SCN analyses are summarized for enhancing stakeholder collaboration in disaster response.

Academically, this research proposes a systematic approach for reliably extracting stakeholder collaboration information from texts and automatically generating SCNs. In practice, the proposed approach largely reduces practitioners' interpretation loads and eases the data collection process. Although the proposed approach achieves automated generation of SCNs, it only considers government agencies as stakeholders. In addition, entities that have agency names but do not refer to stakeholders cannot be distinguished. In the future, the specified dictionary will be enriched by incorporating more types of stakeholders (e.g., nonprofit organizations and private sectors). In addition, more advanced NLP techniques (e.g., semantic role labeling) will be applied to identify entities that do not refer to stakeholders.